\documentclass[twocolumn,aps,superscriptaddress,showpacs]{revtex4}

\usepackage{amssymb}
\usepackage{amsmath}
\usepackage{graphicx}
\usepackage[normalem]{ulem}
\usepackage[dvips]{color}

\begin{document}

\title{Density matrix expansion for the MDI interaction}
\author{Jun Xu}\affiliation{Cyclotron Institute, Texas A\&M University, College
Station, TX 77843-3366, USA}
\author{Che Ming Ko}\affiliation{Cyclotron Institute and Department of Physics and
Astronomy, Texas A\&M University, College Station, TX 77843-3366,
U.S.A.}

\begin{abstract}

By assuming that the isospin- and momentum-dependent MDI interaction
has a form similar to the Gogny-like effective two-body interaction
with a Yukawa finite-range term and the momentum dependence only
originates from the finite-range exchange interaction, we determine
its parameters by comparing the predicted potential energy density
functional in uniform nuclear matter with what has been usually
given and used extensively in transport models for studying isospin
effects in intermediate-energy heavy-ion collisions as well as in
investigating the properties of hot asymmetric nuclear matter and
neutron star matter. We then use the density matrix expansion to
derive from the resulting finite-range exchange interaction an
effective Skyrme-like zero-range interaction with density-dependent
parameters. As an application, we study the transition density and
pressure at the inner edge of neutron star crusts using the
stability conditions derived from the linearized Vlasov equation for
the neutron star matter.

\end{abstract}

\pacs{26.60.-c, 
      21.30.Fe, 
      21.65.-f, 
      97.60.Jd  
      }

\maketitle

\section{Introduction}\label{introduction}

One of the phenomenological nucleon-nucleon interactions that have
been extensively used in transport models to study heavy-ion
collisions is the isospin- and momentum-dependent MDI
interaction~\cite{Das03}. By using this interaction in the
isospin-dependent Boltzmann-Uhling-Uhlenbeck (IBUU) transport model,
an extensive amount of works have been carried out to study
various isospin sensitive observables in intermediate-energy
heavy-ion collisions (for recent reviews see
Refs.~\cite{Chen07,LCK08}). From comparisons of the isospin
diffusion data from NSCL-MSU for $^{124}$Sn+$^{112}$Sn reactions at
$E=50$ MeV/nucleon~\cite{Tsa04} with results from the IBUU model, a
relatively stringent constraint on the density dependence of the
nuclear symmetry energy at subsaturation densities has been
obtained~\cite{Che05a,LiBA05}. The resulting symmetry energy has
further been used to impose constraints on both the parameters in
the Skyrme effective interactions and the neutron skin thickness of
heavy nuclei~\cite{Che05}. The MDI interaction with the constrained
isospin dependence has also been used to study the properties of hot
asymmetric nuclear matter~\cite{Xu08} as well as those of neutron
stars~\cite{Steiner06,Krastev08a,Krastev08b,Wen09,Newton09,bali08},
including the transition density which separates their liquid core
from their inner crust~\cite{XCLM09}. New constraints on the masses
and radii of neutron stars were then obtained from comparing the
resulting crustal fraction of the moment of inertia of neutron stars
with that of the Vela pulsar extracted from its
glitches~\cite{XCLM09}.

Although the momentum-dependent nuclear
mean-field potential or the energy density functional of uniform
nuclear matter was used in the above mentioned applications of the
MDI interaction, the explicit form of the MDI interaction has never
been given in the literatures. It was, however, implicitly
mentioned~\cite{Das03,Gal90} that
this momentum-dependent nucleon mean-field potential was constructed
according to a Gogny-like interaction~\cite{Gog80} consisting of a
zero-range Skyrme-like interaction~\cite{Vau72} and a finite-range
Yukawa interaction as in the MDYI interaction~\cite{Gal90}.
Similar to the Gogny interaction, the momentum
dependence in the nucleon mean-field potential from the MDI
interaction is then from the contribution of the finite-range Yukawa
interaction to the exchange energy density of uniform nuclear
matter. The effect of the finite-range part of the MDI interaction
in nonuniform nuclear matter was in previous applications either
neglected or inconsistently included by using the average
density-gradient terms from the phenomenological Skyrme
interactions.  From comparing the potential energy density
functional obtained from above assumed interaction with that used in
previous applications of the MDI interaction, we are able to find
unique relations among the parameters of this interaction and those
used in the potential energy density functional for uniform nuclear
matter.

As a first step to facility the consistent application of the
finite-range MDI interaction, it is of interest to derive an
effective zero-range interaction with density-dependent parameters
using the density matrix expansion first introduced in
Ref.~\cite{Neg72}. As shown in Ref.~\cite{Spr75} for other
finite-range interactions, corresponding zero-range interactions
derived from the density matrix expansion reproduce reasonably well
in the self-consistent Hartree-Fock approach the binding energies
and radii of nuclei obtained with the original interactions,
particularly if the density matrix expansion is only used for the
exchange energy from the finite-range
interaction~\cite{Neg72,Spr75}. In the present paper, we carry out
such a study as an attempt to include consistently the effect due to
the finite-range part of the MDI interaction. As an application, we
use the resulting interaction to study the transition density and
pressure in neutron stars and compare the results with those from
previous studies using the MDI interaction but with inconsistent
density-gradient coefficients~\cite{XCLM09,Xu10}.

This paper is organized as follows. In Sec.~\ref{model}, we review
the isospin- and momentum-dependent MDI interaction and determine
its underlying nucleon-nucleon (NN) interaction by fitting the
energy density functional used in previous applications of the MDI
interaction. In Sec.~\ref{dme}, the density matrix expansion is then
used to derive from the finite-range exchange interaction a
zero-range effective interaction with density-dependent parameters.
The application of the resulting interaction to study the transition
density and pressure at the inner edge of neutron star crusts is
given in Sec.~\ref{app}, using the stability conditions that are
derived from the linearized Vlasov equation for the neutron star
matter. Finally, the summary is given in Sec.~\ref{sum}. For details
on the determination of the underlying nucleon-nucleon interaction
potential in the MDI interaction and the application of the density
matrix expansion to the exchange energy of nuclear matter, they are
given in Appendices \ref{appa} and \ref{appb}, respectively.

\section{The MDI interaction}\label{model}

In applications of the momentum-dependent MDI interaction for
studying isospin effects in intermediate-energy heavy-ion collisions
as well as the properties of hot asymmetric nuclear matter and
neutron star matter, one usually uses following potential energy
density
\begin{eqnarray}
H(\rho,\delta ) &=&\frac{A_{1}}{2\rho _{0}}\rho^2
+\frac{A_{2}}{2\rho _{0}}\rho^2\delta^2+\frac{B}{\sigma
+1}\frac{\rho ^{\sigma +1}}{\rho
_{0}^{\sigma }}(1-x\delta ^{2})\notag\\
&+&\frac{1}{\rho _{0}}\sum_{\tau ,\tau^{\prime}}C_{\tau ,\tau ^{\prime }}
\int \int d^{3}pd^{3}p^{\prime }\frac{f_{\tau }(\vec{r},\vec{p}%
)f_{\tau ^{\prime }}(\vec{r},\vec{p}^{\prime
})}{1+(\vec{p}-\vec{p}^{\prime })^{2}/\Lambda ^{2}}\notag\\
\label{MDIV}
\end{eqnarray}%
for infinite nuclear matter of density $\rho$ and isospin asymmetry
$\delta=(\rho_n-\rho_p)/\rho$, with $\rho_n$ and $\rho_p$ being,
respectively, the neutron and proton densities. In the above,
$\tau(\tau^\prime)$ is the nucleon isospin;
$f_{\tau}(\vec{r},\vec{p})$ is the nucleon phase-space distribution
function at position $\vec{r}$; and $\rho_0=0.16$ fm$^{-3}$ is the
saturation density of normal nuclear matter. Values of the
parameters $A_1=(A_l+A_u)/2$, $A_2=(A_l-A_u)/2$, $B$, $\sigma$,
$\Lambda$, $C_l=C_{\tau,\tau}$ and $C_u=C_{\tau,-\tau}$ can be found
in Refs.~\cite{Das03,Che05a}. For symmetric nuclear matter, this
interaction gives a binding energy of $-16$ MeV per nucleon and an
incompressibility $K_0$ of $212$ MeV at saturation density.

The parameter $x$ in Eq.~(\ref{MDIV}) is used to model the density
dependence of the symmetry energy. Based on analyses of the isospin
diffusion in intermediate-energy heavy-ion collisions and the
neutron skin thickness of heavy nuclei~\cite{Tsa04,Che05a,LiBA05},
the slope parameter $L=3\rho_0 (\partial E_{\rm
sym}(\rho)/\partial\rho)_{\rho=\rho_0}$ of the symmetry energy has
been constrained to $L=86 \pm 25$ MeV, which corresponds to $-1 < x
< 0$. More recent analyses of experimental data on isospin
diffusion, double $n/p$ ratio, and neutron skin thickness favor,
however, a softer symmetry energy of $L=58 \pm 18$
MeV~\cite{Tsa09,She10,Che10}.

The momentum dependence in the MDI interaction can come from a
number of different origins. Besides the finite-range exchange term
in the nucleon-nucleon interaction, the intrinsic momentum
dependence in the nucleon-nucleon interaction and the effects of
short-range nucleon-nucleon correlations~\cite{Pan05} can also
contribute to its momentum dependence. For simplicity, we assume in
the present study that all of the momentum dependence in the MDI
interaction comes from the finite-range exchange term. In this case,
the explicit form for the MDI interaction can be obtained from the
energy density given in Eq.~(\ref{MDIV}) by assuming that the
interaction potential between two nucleons located at $\vec{r}_1$
and $\vec{r}_2$ has a form similar to the Gogny
interaction~\cite{Gog80,Das03} but with its Gaussian form in the
finite-range term replaced by a Yukawa form, that is
\begin{eqnarray}
v(\vec{r}_1,\vec{r}_2) &=& \frac{1}{6}t_3(1+x_3 P_\sigma)
\rho^\alpha\left(\frac{\vec{r}_1+\vec{r}_2}{2}\right)
\delta(\vec{r}_1-\vec{r}_2) \notag\\
&+& (W+B P_\sigma - H P_\tau - M P_\sigma P_\tau) \frac{e^{-\mu
|\vec{r}_1-\vec{r}_2}|}{|\vec{r}_1-\vec{r}_2|}.\notag\\
\label{MDIv}
\end{eqnarray}
In the above equation, the first term is the density-dependent
zero-range interaction which can be considered as an effective
three-body interaction, whereas the second term is the
density-independent finite-range interaction. In terms of this
nucleon-nucleon interaction, the total potential energy of a nuclear
system can be calculated from
\begin{equation}\label{vV}
E = \frac{1}{2} \sum_{i,j} <ij|v (1-P_r P_\sigma P_\tau)|ij>,
\end{equation}
where
\begin{equation}
|i> = |i_r i_\sigma i_\tau>
\end{equation}
is the quantum state of nucleon $i$ with the spatial state $i_r$,
the spin state $i_\sigma$, and the isospin state $i_\tau$, and
$P_r$, $P_\sigma$ and $P_\tau$ are, respectively, the space, spin
and isospin exchange operators.

As shown in Appendix \ref{appa}, where the detailed derivation of
Eq.~(\ref{MDIV}) from Eq.~(\ref{MDIv}) is given, the first and
second terms in Eq.~(\ref{MDIV}) come from the direct contribution
(the first term of Eq.~(\ref{vV})) of the finite-range term in
the NN interaction. The third term in Eq.~(\ref{MDIV}) is simply from the contribution
of the zero-range term. Although different density-dependent
zero-range terms can lead to different symmetry
energies/potentials~\cite{XuC10}, we use in the present study the
one in the standard Skyrme interaction~\cite{Cha97}. The
momentum-dependent terms in Eq.~(\ref{MDIV}) come from the exchange
contribution (the second term of Eq.~(\ref{vV})) of the
finite-range interaction. Comparing the resulting energy density functional
with Eq.~(\ref{MDIV}) leads to following unique relations between
the 8 parameters in the NN interaction and those in the energy
density functional of Eq.~(\ref{MDIV}):
\begin{eqnarray}
t_3 &=& \frac{16 B}{(\sigma+1)\rho_0^\sigma},\label{t3}\\
x_3 &=& \frac{3x-1}{2},\label{x3}\\
\alpha &=& \sigma - 1,\label{alpha}\\
\mu &=& \Lambda, \label{lambda} \\
W &=& \frac{\Lambda^2}{3 \pi \rho_0} (A_1-A_2+C_l-C_u),\label{w}\\
B &=& \frac{\Lambda^2}{6 \pi \rho_0} (-A_1+A_2-4C_l+4C_u),\label{b}\\
H &=& \frac{\Lambda^2}{3 \pi \rho_0} (-2A_2-C_u),\label{h}\\
M &=& \frac{\Lambda^2}{3 \pi \rho_0} (A_2+2C_u).\label{m}
\end{eqnarray}
It is seen that the symmetry energy parameter $x$ is related to the
coefficient of the spin-exchange term in the density-dependent
zero-range interaction, whereas the width $\Lambda$ in the momentum
dependence is related to the range of the finite-range interaction.

In the original MDI interaction~\cite{Das03}, the symmetry energy
parameter $x$ can only have the value of $0$ or $1$. To get a larger
range of density dependence for the symmetry energy while fixing the
value of $E_{\rm sym}(\rho_0)=30.5$ MeV, the parameters $A_l$ and
$A_u$ have been expressed as~\cite{Che05a}
\begin{equation}
A_l = -120.57 + x \frac{2B}{ \sigma + 1},~ A_u = -95.98 - x
\frac{2B}{\sigma + 1},
\end{equation}
which results in the $x$ dependence of $A_2$. Therefore, with this constraint the $x$
dependence also appears in the finite-range direct interaction.

\section{The density Matrix Expansion}
\label{dme}

Although the contribution of a finite-range interaction to the
direct energy density of nuclear matter can be treated exactly, it
is numerically challenging to evaluate its contribution to the
exchange energy density. The latter can be, however, approximated by
that from a Skyrme-like zero-range interaction using the
density-matrix expansion of Ref.~\cite{Neg72}. As shown in Appendix
\ref{appb}, the resulting exchange energy density from the
finite-range interaction at position $\vec{r}$ in a nuclear system
can be expressed in terms of densities $\rho_n$ and $\rho_p$ as well
as kinetic energy densities $\tau_n$ and $\tau_p$ as
\begin{eqnarray}\label{hfh}
H_{SL}^E(\vec{r}) &=& A[\rho_n(\vec{r}),\rho_p(\vec{r})] \notag\\
&+& B[\rho_n(\vec{r}),\rho_p(\vec{r})] \tau_n(\vec{r}) +
B[\rho_p(\vec{r}),\rho_n(\vec{r})] \tau_p(\vec{r}) \notag\\
&+& C[\rho_n(\vec{r}),\rho_p(\vec{r})] |\nabla \rho_n(\vec{r})|^2
\notag\\
&+& C[\rho_p(\vec{r}),\rho_n(\vec{r})] |\nabla \rho_p(\vec{r})|^2
\notag\\
&+& D[\rho_n(\vec{r}),\rho_p(\vec{r})] \nabla \rho_n(\vec{r}) \cdot
\nabla \rho_p(\vec{r}),
\end{eqnarray}
where
\begin{eqnarray}
A(\rho_n,\rho_p) &=& V_{NM}(\rho_n,\rho_p)  -
\frac{3}{5}(3\pi^2)^{2/3}[\rho_n^{5/3} B(\rho_n,\rho_p) \notag\\
&+& \rho_p^{5/3} B(\rho_p,\rho_n)],\\
B(\rho_n,\rho_p) &=& -2[\rho_n V^L(\rho_n) + \rho_p
V^U(\rho_p,\rho_n)],\\
C(\rho_n,\rho_p) &=& -\frac{\partial
F(\rho_n,\rho_p)}{\partial \rho_n},\\
D(\rho_n,\rho_p) &=& - \frac{\partial F(\rho_n,\rho_p)}{\partial
\rho_p} - \frac{\partial
F(\rho_p,\rho_n)}{\partial \rho_n}.\notag\\
\end{eqnarray}
In above equations, $V_{NM}(\rho_n,\rho_p)$ is the potential energy
density of infinite nuclear matter from the finite-range
exchange interaction (i.e. the momentum-dependent terms in Eq.~(\ref{MDIV})) and is
given by
\begin{eqnarray}
V_{NM}(\rho_n,\rho_p) &=&  \rho_n^2 V_{NM}^{L}(\rho_n) + \rho_p^2 V_{NM}^{L}(\rho_p)\notag\\
&+& 2\rho_n \rho_p V_{NM}^{U} (\rho_n,\rho_p),
\end{eqnarray}
with
\begin{eqnarray}
V_{NM}^{L}(\rho_\tau) &=& \frac{1}{2}
\left(M+\frac{H}{2}-B-\frac{W}{2}\right)
\notag\\
&\times& \int d^3s \rho_{SL}^2(k_\tau s)\frac{e^{-\mu s}}{s},
\\
V_{NM}^{U}\left(\rho_\tau,\rho_{\tau^\prime}\right) &=& \frac{1}{2}
\left(M+\frac{H}{2}\right) \notag\\
&\times&\int d^3s \rho_{SL}^{}(k_\tau
s)\rho_{SL}^{}(k_{\tau^\prime}s)\frac{e^{-\mu s}}{s}.
\end{eqnarray}
Other terms are defined as
\begin{eqnarray}
V^L(\rho_\tau) &=& \frac{1}{2} \left(M+\frac{H}{2}-B-\frac{W}{2}\right) \notag\\
&\times& \int d^3s s^2 \rho_{SL}^{}(k_\tau s) g(k_\tau s) \frac{e^{-\mu s}}{s},\\
V^U(\rho_\tau,\rho_{\tau^\prime}) &=& \frac{1}{2}
\left(M+\frac{H}{2}\right)
\notag\\
&\times& \int d^3s s^2 \rho_{SL}^{}(k_\tau s) g(k_{\tau^\prime} s)
\frac{e^{-\mu s}}{s},
\end{eqnarray}
where
\begin{eqnarray}
\rho_{SL}^{}(k_\tau s) &=& \frac{3}{k_\tau s}j_1(k_\tau s), \\
g(k_\tau s) &=& \frac{35}{2(k_\tau s)^3}j_3(k_\tau s),
\end{eqnarray}
with $j_1$ and $j_3$ being, respectively, the first- and third-order
spherical bessel functions and $k_\tau=(3\pi^2\rho_\tau)^{1/3}$ is
the Fermi momentum. For the function $F(\rho_n,\rho_p)$, it is
defined as
\begin{eqnarray}
F(\rho_n,\rho_p) = \frac{1}{2}V^L(\rho_n)\rho_n
+\frac{1}{2}V^U(\rho_p,\rho_n)\rho_p.
\end{eqnarray}

\begin{figure}[h]
\begin{minipage}[b]{1.0\linewidth}
\hspace{-1cm}\includegraphics[scale=0.5]{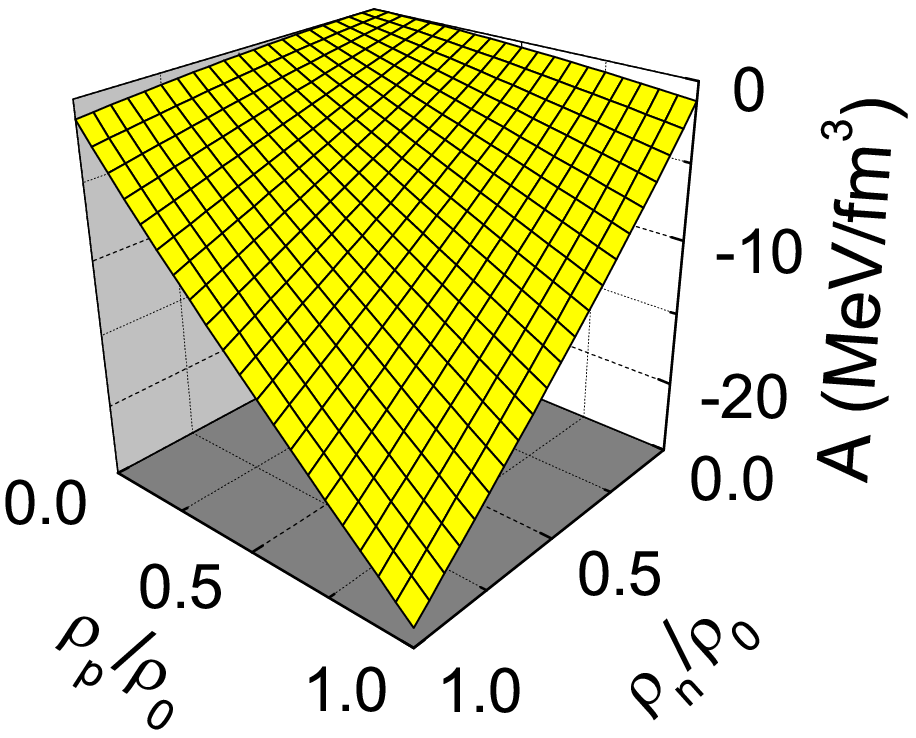}\includegraphics[scale=0.5]{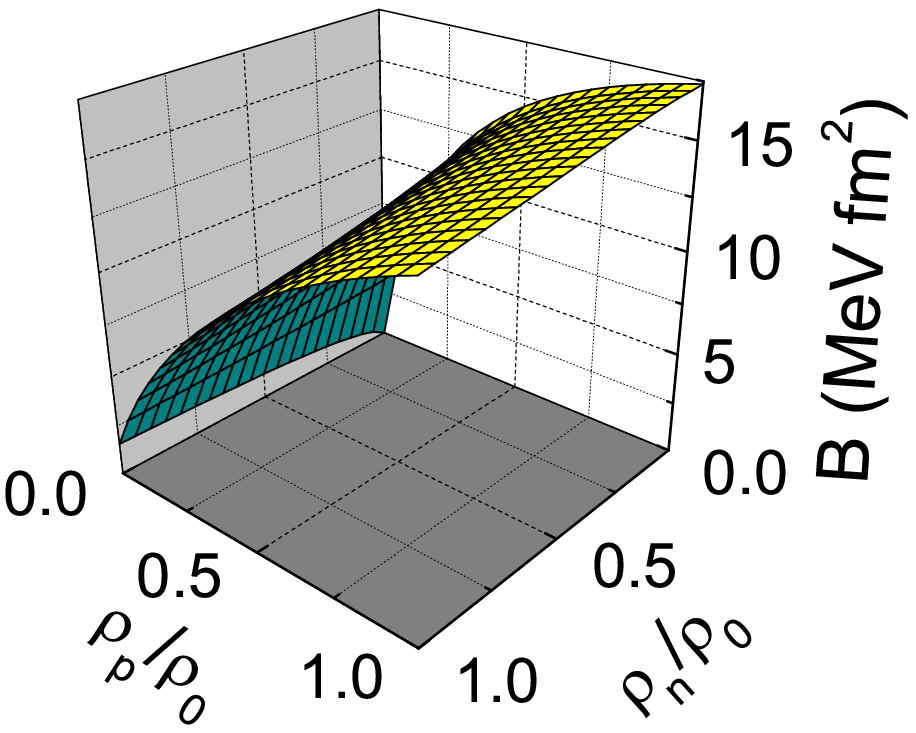}
\end{minipage}
\begin{minipage}[b]{1.0\linewidth}
\hspace{-1cm}\includegraphics[scale=0.5]{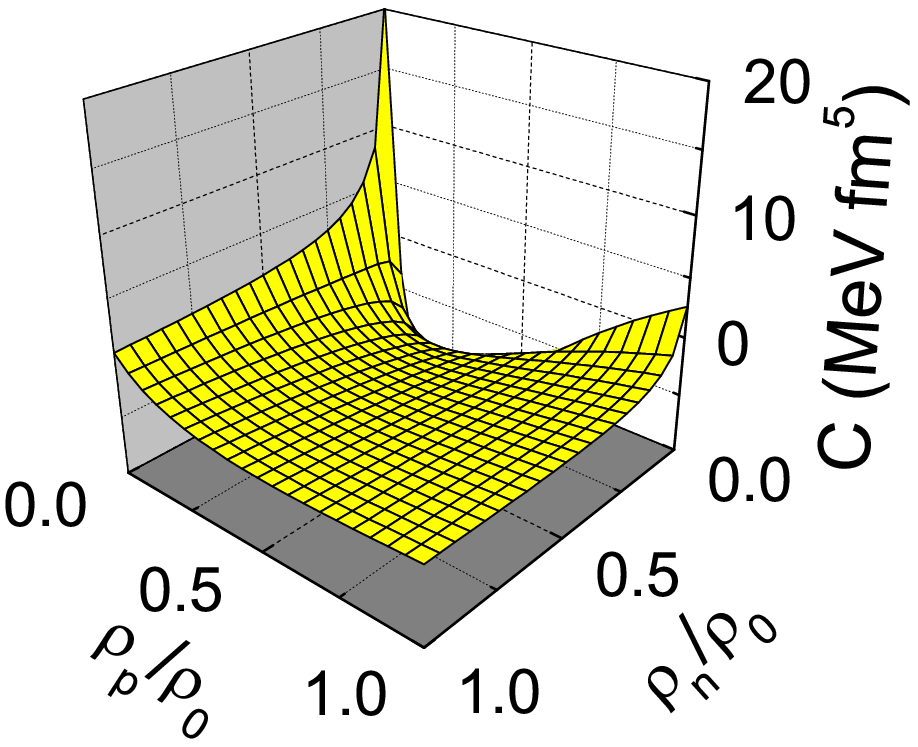}\includegraphics[scale=0.5]{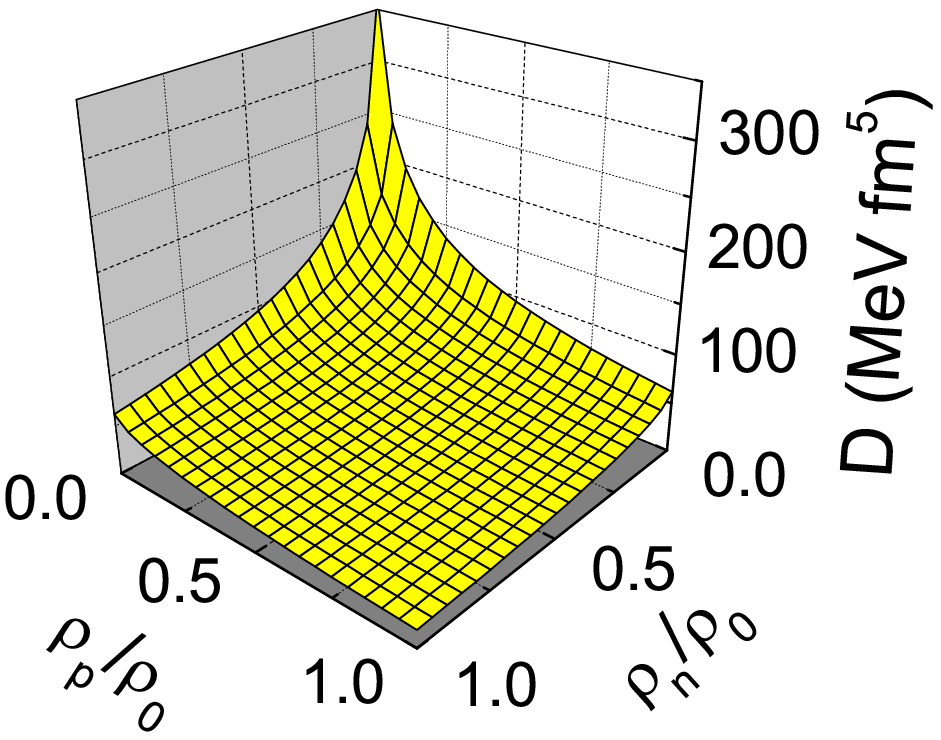}
\end{minipage}
\caption{(Color online) Dependence of $A(\rho_n,\rho_p)$,
$B(\rho_n,\rho_p)$, $C(\rho_n,\rho_p)$ and $D(\rho_n,\rho_p)$ on
$\rho_n$ and $\rho_p$ at subsaturation densities. }\label{abcd}
\end{figure}

In Fig.~\ref{abcd}, we show the dependence of $A(\rho_n,\rho_p)$,
$B(\rho_n,\rho_p)$, $C(\rho_n,\rho_p)$ and $D(\rho_n,\rho_p)$ on
$\rho_n$ and $\rho_p$ at subsaturation densities. It is seen that
$A(\rho_n,\rho_p)$ and $D(\rho_n,\rho_p)$ are symmetric in $\rho_n$
and $\rho_p$, while $B(\rho_n,\rho_p)$ and $C(\rho_n,\rho_p)$ are
not. Furthermore, the density dependence of $B(\rho_n,\rho_p)$,
$C(\rho_n,\rho_p)$ and $D(\rho_n,\rho_p)$ are strong at low
densities but weak near the saturation density. As these functions
are from the exchange contribution of the finite-range interaction,
they are independent of $x$.

\section{Application: transition density and pressure in neutron stars}
\label{app}

The explicit form of the finite-range NN interaction and the
Skyrme-like zero-range interaction with density-dependent parameters
derived in previous sections make it convenient to use the MDI
interaction for a wider range of studies. In this Section, we
discuss its application in studying the transition density and
pressure at the inner edge of neutron star crusts based on the
stability conditions that are derived from the linearized Vlasov
equation for the neutron star matter.

\subsection{The single-particle Hamiltonian}

The single-particle Hamiltonian for a nucleon can be obtained from
minimizing the total energy of a nuclear system with respect to its
wave function. For a neutron, it is given by
\begin{eqnarray}\label{hn}
h_n &=& - \nabla \cdot \left( \frac{\hbar^2}{2m_n^\star} \nabla
\right) + U_n(\rho_n,\rho_p)
\end{eqnarray}
with the neutron effective mass
\begin{equation}
\frac{\hbar^2}{2 m_n^\star} = \frac{\hbar^2}{2 m} +
B(\rho_n,\rho_p),
\end{equation}
and the potential
\begin{equation}
U_n(\rho_n,\rho_p) = U_n^\rho(\rho_n,\rho_p) + U_n^D(\rho_n,\rho_p)
+ U_n^E(\rho_n,\rho_p),
\end{equation}
where
\begin{equation}
U_n^\rho(\rho_n,\rho_p) = \frac{\partial
H_0(\rho_n,\rho_p)}{\partial \rho_n}
\end{equation}
is the contribution from the the zero-range interaction with
\begin{equation}\label{hrho}
H_0(\rho_n,\rho_p) = \frac{1}{24} t_3 \rho^\alpha
[(2+x_3)\rho^2-(2x_3+1)(\rho_n^2+\rho_p^2)]
\end{equation}
being the corresponding energy density, and
\begin{eqnarray}
U_n^D(\rho_n,\rho_p) &=& (W+\frac{B}{2}-H-\frac{M}{2}) \int d^3 r^\prime
\rho_n(\vec{r}^\prime) \frac{e^{-\mu
|\vec{r}-\vec{r}^\prime}|}{|\vec{r}-\vec{r}^\prime|} \notag\\
&+& (W+\frac{B}{2}) \int d^3 r^\prime \rho_p(\vec{r}^\prime)
\frac{e^{-\mu |\vec{r}-\vec{r}^\prime}|}{|\vec{r}-\vec{r}^\prime|},
\end{eqnarray}
\begin{eqnarray}
U_n^E(\rho_n,\rho_p) &=& \frac{\partial
A(\rho_n,\rho_p)}{\partial \rho_n} + \frac{\partial
B(\rho_n,\rho_p)}{\partial \rho_n} \tau_n \notag\\
&+& \frac{\partial B(\rho_p,\rho_n)}{\partial \rho_n} \tau_p -
\frac{\partial C(\rho_n,\rho_p)}{\partial \rho_n} (\nabla \rho_n)^2
\notag\\
&+& \left( \frac{\partial C(\rho_p,\rho_n)}{\partial
\rho_n}-\frac{\partial D(\rho_n,\rho_p)}{\partial \rho_p}\right)
(\nabla \rho_p)^2 \notag\\
&-& 2\frac{\partial C(\rho_n,\rho_p)}{\partial \rho_p} \nabla \rho_n
\cdot \nabla \rho_p - 2 C(\rho_n,\rho_p) \nabla^2 \rho_n \notag\\
&-& D(\rho_n,\rho_p) \nabla^2 \rho_p
\end{eqnarray}
are the direct and exchange contributions from the finite-range
interaction, respectively. Expressing the kinetic energy density in
the exchange potential in terms of the density via the extended
Thomas-Fermi (ETF) approximation~\cite{Jen76,Bra85}
\begin{equation}\label{etf}
\tau_q = a \rho_q^{5/3} + b \frac{(\nabla
\rho_q)^2}{\rho_q} + c \nabla^2 \rho_q,~q=n,p,
\end{equation}
where $a = \frac{3}{5}(3\pi^2)^{2/3}$, $b=1/36$ and $c=1/3$, the
neutron potential can then be written as
\begin{eqnarray}
U_n &=&  U_n^0 + U_n^D + U_n^\nabla,
\end{eqnarray}
where
\begin{eqnarray}
U_n^0(\rho_n,\rho_p) &=& U_n^\rho(\rho_n,\rho_p) + \frac{\partial
A(\rho_n,\rho_p)}{\partial \rho_n} \notag\\
&+& a \frac{\partial B(\rho_n,\rho_p)}{\partial \rho_n}
\rho_n^{5/3} + a \frac{\partial B(\rho_p,\rho_n)}{\partial
\rho_n} \rho_p^{5/3}, \\
U_n^\nabla(\rho_n,\rho_p) &=& G_n^{nn1} (\nabla
\rho_n)^2 + G_n^{pp1} (\nabla \rho_p)^2 \notag\\
&+& G_n^{np} \nabla \rho_n \cdot \nabla \rho_p + G_n^{nn2} \nabla^2
\rho_n + G_n^{pp2} \nabla^2 \rho_p,\notag\\
\end{eqnarray}
with
\begin{eqnarray}
G_n^{nn1}(\rho_n,\rho_p) &=& -\frac{\partial
C(\rho_n,\rho_p)}{\partial \rho_n} + \frac{b}{\rho_n} \frac{\partial
B(\rho_n,\rho_p)}{\partial \rho_n}, \\
G_n^{pp1}(\rho_n,\rho_p) &=& \frac{\partial
C(\rho_p,\rho_n)}{\partial \rho_n} - \frac{\partial
D(\rho_n,\rho_p)}{\partial \rho_p} \notag\\
&+& \frac{b}{\rho_p} \frac{\partial
B(\rho_p,\rho_n)}{\partial \rho_n}, \\
G_n^{np}(\rho_n,\rho_p) &=& -2\frac{\partial
C(\rho_n,\rho_p)}{\partial \rho_p},\\
G_n^{nn2}(\rho_n,\rho_p) &=& -2 C(\rho_n,\rho_p) + c
\frac{\partial
B(\rho_n,\rho_p)}{\partial \rho_n}, \\
G_n^{pp2}(\rho_n,\rho_p) &=& -D(\rho_n,\rho_p) + c
\frac{\partial
B(\rho_p,\rho_n)}{\partial \rho_n}.
\end{eqnarray}
Similarly, the proton single-particle Hamiltonian is
\begin{eqnarray}
h_p = h_n(n \leftrightarrow p) + U_p^{CD} + U_p^{CE},
\end{eqnarray}
where the first term $h_n(n \leftrightarrow p)$ is due to the
nuclear interaction and is obtained from Eq.~(\ref{hn}) by
interchanging the neutron and proton symbols; the second term
$U_p^{CD}$ is the direct Coulomb potential given by
\begin{equation}
U_p^{CD} = e^2 \int \frac{\rho_p(\vec{r}^\prime) -
\rho_e(\vec{r}^\prime)}{|\vec{r}-\vec{r}^\prime|} d^3 r^\prime;
\end{equation}
and the last term $U_p^{CE}$ is the exchange Coulomb potential which
in the first order of the density matrix expansion~\cite{Neg72} is
given by
\begin{equation}
U_p^{CE} = - e^2 \left(\frac{3}{\pi}\right)^{1/3} \rho_p^{1/3}.
\end{equation}

\subsection{The Vlasov equation for the neutron star matter}

The Vlasov equation has been widely used in studying the collective
density fluctuation and spinodal instability in nuclear matter (See
Ref.~\cite{Cho04} for a recent review). For a $\beta$-stable and
charge neutral $npe$ matter, the Vlasov equation can be written as
\begin{equation}\label{vlasov}
\frac{\partial f_q(\vec{r},\vec{p},t)}{\partial t} + \vec{v}_q \cdot
\nabla_{\vec{r}} f_q(\vec{r},\vec{p},t) - \nabla_{\vec{r}} U_q \cdot
\nabla_{\vec{p}} f_q(\vec{r},\vec{p},t) = 0
\end{equation}
in terms of the Wigner function for particle type $q=n,p,e$
\begin{eqnarray}
f_q(\vec{r},\vec{p},t) &=& \frac{1}{(2\pi)^3} \sum_i \int
\phi_{qi}\left(\vec{r}-\frac{\vec{s}}{2},t\right)\notag\\
&\times&\phi_{qi}^\star\left(\vec{r}+\frac{\vec{s}}{2},t\right)
e^{i\vec{p} \cdot \vec{s}} d^3 s,
\end{eqnarray}
where $\phi_{qi}$ is the wave function of i-th particle of type $q$.
In Eq.~(\ref{vlasov}), $\vec{v}_q
=\vec{p}/{(m_q^\star}^2+p^2)^{1/2}$ denotes the particle velocity
and $m_e^\star=m_e$.

To study the density fluctuation due to a collective mode with
frequency $\omega$ and wavevector $\vec{k}$ in the nuclear matter,
we follow the standard procedure~\cite{Cho04} by writing
\begin{equation}
f_q(\vec{r},\vec{p},t) = f_q^0(\vec{p}) + \delta f_q
(\vec{r},\vec{p},t)
\end{equation}
with
\begin{equation}
\delta f_q (\vec{r},\vec{p},t) = \delta \tilde{f_q} (\vec{p}) e^{-i\omega
t+i \vec{k}\cdot\vec{r}}.
\end{equation}
As the momentum dependence of the Wigner function is mainly from
$f_q^0(\vec{p})$, we have $\nabla_{\vec{p}}f_q(\vec{r},\vec{p},t)\approx \nabla_{\vec{p}}
f^0_q(\vec{p})$ and can thus rewrite the Vlasov equation (Eq.~(\ref{vlasov})) as
\begin{eqnarray}\label{vlasov1}
-i\omega \delta f_q &+& \vec{v}_q \cdot (i\vec{k}) \delta f_q \notag\\
&-&\frac{\partial f_q^0}{\partial \epsilon_q}
\nabla_{\vec{p}}\epsilon_q \cdot \left( \sum_{q^\prime} \frac{\delta
U_q}{\delta \rho_{q^\prime}} \nabla_{\vec{r}}
\rho_{q^\prime}\right)=0,\notag\\
\end{eqnarray}
where $\epsilon_q$ is the single-particle energy. Using
$\nabla_{\vec{p}}\epsilon_q = \vec{v}_q$ and writing
$\rho_q(\vec{r},t) = \rho_q^0 + \delta\rho_q (\vec{r},t)$ with
\begin{equation}\label{vlasov2}
\delta \rho_q(\vec{r},t) = \frac{2}{(2\pi)^3} \int \delta
f_q(\vec{r},\vec{p},t) d^3p,
\end{equation}
which has same time and spatial dependence as $\delta
f_q(\vec{r},\vec{p},t)$, Eq.~(\ref{vlasov1}) can be rewritten as
\begin{equation}
\delta f_q = - \frac{\partial f_q^0}{\partial \epsilon_q} \left(
\sum_{q^\prime} \frac{\delta U_q}{\delta \rho_{q^\prime}} \delta
\rho_{q^\prime}\right) \frac{\vec{k} \cdot \vec{v}_q}{\omega -
\vec{k} \cdot \vec{v}_q}.
\end{equation}
By substituting the above equation into Eq.~(\ref{vlasov2}), we
obtain in the low-temperature limit
\begin{eqnarray}\label{drho}
\delta \rho_q &\approx&
\frac{1}{2\pi^2}\int_0^{p_q^F}\left(-\frac{\partial f_q^0}{\partial
\epsilon_q}\right) p^2 dp \int_{-1}^{1} \frac{\cos\theta
d(\cos\theta)}{s_q-\cos\theta} \notag\\
&\times&\left( \sum_{q^\prime} \frac{\delta U_q}{\delta
\rho_{q^\prime}} \delta \rho_{q^\prime}\right),
\end{eqnarray}
with $s_q=\omega/k v_q^F$ and $v_q^F =p_q^F/
({m_q^\star}^2+{p_q^F}^2)^{1/2}$ being the Fermi velocity. Carrying
out the angular integration leads to the usual Lindhard function
\begin{eqnarray}
L_q = \int_{-1}^{1} \frac{\cos\theta d(\cos\theta)}{s_q-\cos\theta}
= -2 + s_q \ln \left( \frac{s_q+1}{s_q-1}\right).
\end{eqnarray}
In the low-temperature limit, the momentum integration can be
approximately evaluated as
\begin{eqnarray}\label{xq}
X_q &=& \frac{1}{2\pi^2} \int_0^{p_q^F} \left(-\frac{\partial
f_q^0}{\partial\epsilon_q}\right) p^2 dp \notag\\
&\approx& \frac{p_q^F m_q^\star}{2\pi^2} \left[ 1 - \frac{\pi^2}{24}
\left( \frac{T}{\epsilon_q^F}\right)^2\right]
\end{eqnarray}
for $q=n,p$, with $\epsilon_q^F \approx{p_q^F}^2/2m_q^\star$ and
\begin{equation}\label{xe}
X_e \approx \frac{\mu_e^2}{2\pi^2} \left[ 1 + \frac{\pi^2}{3} \left(
\frac{T}{\mu_e}\right)^2\right],
\end{equation}
for electrons, where $\mu_e \approx p_e^F$ is the electron chemical
potential.

For the factor $\sum_{q^\prime} \frac{\delta U_q}{\delta
\rho_{q^\prime}} \delta \rho_{q^\prime}$ in Eq.~(\ref{drho}), there
are the local contribution $\delta U_n^0$, the direct contribution
$\delta U_n^D$ from the finite-range interaction, and the
gradient contribution $\delta U_n^\nabla$
from the finite-range exchange interaction using the density matrix
expansion. In the case of neutrons, they are given by
\begin{eqnarray}
\delta U_n^0 &=& \frac{\partial U_n^0}{\partial \rho_n} \delta
\rho_n
+ \frac{\partial U_n^0}{\partial \rho_p} \delta \rho_p,\\
\delta U_n^D &=& (W+\frac{B}{2}-H-\frac{M}{2}) \frac{4 \pi}{k^2 + \mu^2}
\delta \rho_n \notag \\
&+& (W+\frac{B}{2}) \frac{4 \pi}{k^2 + \mu^2} \delta \rho_p,\\
\delta U_n^\nabla &=& -k^2(G_n^{nn2} \delta \rho_n + G_n^{pp2}
\delta \rho_p),
\end{eqnarray}
after neglecting higher-order terms in $\delta \rho_q$. In the
long-wavelength case, the finite-range direct contribution can be
rewritten as
\begin{equation}
\delta U_n^D \approx \frac{\partial U_n^{D0}}{\partial \rho_n} \delta \rho_n
+ \frac{\partial U_n^{D0}}{\partial \rho_p} \delta \rho_p
- k^2 (G_n^{nnD} \delta \rho_n + G_n^{ppD} \delta \rho_p)
\end{equation}
with
\begin{eqnarray}
\frac{\partial U_n^{D0}}{\partial \rho_n} &=&
\frac{4 \pi}{\mu^2}(W+\frac{B}{2}-H-\frac{M}{2}), \\
\frac{\partial U_n^{D0}}{\partial \rho_p} &=&
\frac{4 \pi}{\mu^2}(W+\frac{B}{2}), \\
G_n^{nnD} &=& \frac{4 \pi}{\mu^4} (W+\frac{B}{2}-H-\frac{M}{2}), \\
G_n^{ppD} &=& \frac{4 \pi}{\mu^4}(W+\frac{B}{2}).
\end{eqnarray}
For protons, there are additional direct and exchange Coulomb
contributions given, respectively, by
\begin{eqnarray}
\delta U_p^{CD} &=& \frac{4\pi e^2}{k^2} (\delta \rho_p - \delta
\rho_e),\\
\delta U_p^{CE} &=& -\frac{1}{3} e^2 \left( \frac{3}{\pi}
\right)^{1/3} \rho_p^{-2/3} \delta \rho_p.
\end{eqnarray}
For electrons, there are only direct and exchange Coulomb
contributions to $\sum_{q^\prime} \frac{\delta U_q}{\delta
\rho_{q^\prime}} \delta \rho_{q^\prime}$.

After linearizing the Vlasov equation, the collective density
fluctuation $\delta \rho_q$ then satisfies the following equation
\begin{equation}\label{eqdrho}
C^f(\delta \rho_n, \delta \rho_p, \delta \rho_e)^T=0,
\end{equation}
with
\begin{widetext}
\begin{eqnarray}\label{matrix}
C^f &=&
\left(\begin{array}{ccc}
X_nL_n(\frac{\partial U_n^0}{\partial \rho_n}+\frac{\partial U_n^{D0}}{\partial \rho_n})  -1 & X_nL_n (\frac{\partial U_n^0}{\partial \rho_p}+\frac{\partial U_n^{D0}}{\partial \rho_p}) & 0\\
X_pL_p(\frac{\partial U_p^0}{\partial \rho_n}+\frac{\partial U_p^{D0}}{\partial \rho_n})  & X_pL_p(\frac{\partial U_p^0}{\partial \rho_p}+\frac{\partial U_p^{D0}}{\partial \rho_p}) -1  & 0\\
0 & 0 & -1\\
\end{array}
\right)\notag\\
&-& k^2 \left(
\begin{array}{ccc}
X_nL_n(G_n^{nnD}+G_n^{nn2}) & X_nL_n(G_n^{ppD}+G_n^{pp2}) & 0\\
X_pL_p(G_p^{nnD}+G_p^{nn2}) & X_pL_p(G_p^{ppD}+G_p^{pp2}) & 0\\
0 & 0 & 0\\
\end{array}
\right)+ \left(
\begin{array}{ccc}
0 & 0 & 0\\
0 & X_pL_p\left(\frac{4\pi e^2}{k^2}+\frac{\delta U_p^{CE}}{\delta \rho_p} \right) & -\frac{4\pi e^2}{k^2}X_pL_p\\
0 & -\frac{4\pi e^2}{k^2}X_eL_e & X_eL_e\left(\frac{4\pi e^2}{k^2}+\frac{\delta U_e^{CE}}{\delta \rho_e}\right)\\
\end{array}
\right).\notag\\
\end{eqnarray}
\end{widetext}
We note that the three terms in the above equation are due to,
respectively, the bulk, the density-gradient, and the Coulomb
contribution.

\subsection{The transition density in neutron stars}

Non-trivial solutions of Eq.~(\ref{eqdrho}) are obtained when
$|C^f|=0$, which also determines the dispersion relation $\omega(k)$
of the collective density fluctuation. The transition density in a
neutron star is the density at which the collective density
fluctuation would grow exponentially, resulting in the instability
of the neutron star matter, and this happens when the frequency
$\omega$ becomes imaginary. To determine the condition for this to
occur, we let $s_q = -i\nu_q$ $(\nu_q>0)$ and rewrite the Lindhard
function as $L_q = -2 + 2\nu_q \arctan (1/\nu_q)$. Since the values
of $L_q$ are $-2<L_q<0$, the critical values $L_n=L_p=L_e=-2$,
corresponding to $\nu_q=0$, then determine the spinodal boundary of
the system when they are substituted into $|C^f|=0$. Expanding the
nucleon and electron densities at low temperatures according to
\begin{eqnarray}
\rho_e &\approx&
\frac{\mu_e^3}{3\pi^2}\left[1+\pi^2\left(\frac{T}{\mu_e}\right)^2\right],\\
\rho_q &\approx& \frac{
(2m_q^\star\epsilon_q^F)^{3/2}}{3\pi^2}\left[ 1 + \frac{\pi^2}{8}
\left(\frac{T}{\epsilon_q^F} \right)^2\right], ~ q=n,p
\end{eqnarray}
we obtain from Eqs.~(\ref{xq}) and (\ref{xe}) the following
relations:
\begin{eqnarray}
1/2X_e &=& \partial \mu_e/\partial \rho_e,\\
1/2X_q &=& \partial \epsilon_q^F/ \partial \rho_q.
\end{eqnarray}
The bulk contribution in Eq.~(\ref{matrix}) in the low-temperature
limit can thus be rewritten as
\begin{eqnarray}
-8\left(
\begin{array}{ccc}
X_n (\partial \mu_n/\partial \rho_n) &  X_n (\partial \mu_n/\partial \rho_p)   & 0\\
X_p(\partial \mu_p/\partial \rho_n)  & X_p(\partial \mu_p/\partial \rho_p)   & 0\\
0 & 0 & X_e(\partial \mu_e/\partial \rho_e)\\
\end{array}
\right).
\end{eqnarray}
Denoting the effective density-gradient coefficients as
\begin{eqnarray}
D_{nn} &=& - G_n^{nnD} - G_n^{nn2},\\
D_{np} &=& - G_n^{ppD} - G_n^{pp2},\\
D_{pn} &=& - G_p^{nnD} - G_p^{nn2},\\
D_{pp} &=& - G_p^{ppD} - G_p^{pp2},
\end{eqnarray}
we then obtain same expressions for determining the spinodal
boundary of $npe$ matter in the so-called dynamical
approach~\cite{BBP71,Pet95b,Oya07,XCLM09}, if the Coulomb exchange
terms are neglected. In particular, in the long-wavelength limit,
the spinodal boundary is determined by the vanishing point of
\begin{equation}\label{Vdyn}
V_{\rm dyn}(k) =\frac{|C^f|_{L_q=-2}}{X_nX_qX_e(\partial \mu_n /
\partial \rho_n)}\approx V_0 + \beta k^2 + \frac{4 \pi e^2}{k^2 +
k^2_{TF}},
\end{equation}
where
\begin{eqnarray}
V_0 &=& \frac{\partial \mu_p}{\partial \rho_p} - \frac{(\partial
\mu_p / \partial \rho_n)^2}{\partial \mu_n / \partial \rho_n} +
\frac{\delta U_p^{CE}}{\delta \rho_p},
\label{v0}\\
\beta &=& D_{pp} + (D_{np} +D_{pn})\zeta + D_{nn} \zeta^2,\\
\zeta &=& -\frac{\partial \mu_p / \partial \rho_n}{\partial \mu_n / \partial \rho_n},\\
k^2_{TF} &=& \frac{4 \pi e^2}{(\partial \mu_e /\partial
\rho_e+\delta U_e^{CE}/\delta \rho_e)}.
\end{eqnarray}
To find the upper limit of the spinodal boundary for all possible
wavevectors, we first find the minimum value of $V_{\rm dyn}$ given
by
\begin{equation}\label{Vdynmin}
V_{\rm dyn} = V_0 + 2 (4 \pi e^2 \beta)^{1/2} - \beta k^2_{TF},
\end{equation}
which occurs when $k^2 = (4\pi e^2/\beta)^{1/2}-k_{TF}^2$. The
density that makes Eq.~(\ref{Vdynmin}) vanish then determines the
spinodal boundary in the neutron star matter or the transition
density at the inner edge of neutron star crusts.

\begin{figure}[h]
\centerline{\includegraphics[scale=0.8]{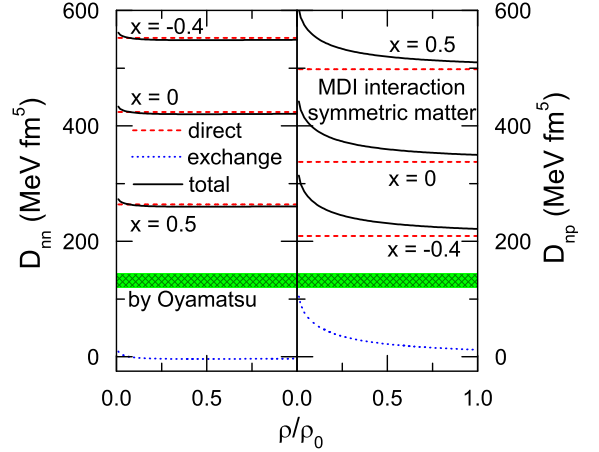}} \caption{(Color
online) Density dependence of effective density-gradient
coefficients from different values of $x$ for symmetric nuclear
matter. The finite-range direct and exchange contributions are
compared, and the value extracted by Oyamatsu~\cite{Oya03} is also
shown. } \label{DD}
\end{figure}

In Fig.~\ref{DD}, we show the density dependence of effective
density-gradient coefficients from different values of $x$ for
symmetric nuclear matter. It is seen that the contribution from the
finite-range direct interaction is much larger than that from the
finite-range exchange interaction. Also, $D_{nn}$ decreases while
$D_{np}$ increases with increasing value of $x$. Furthermore,
$D_{nn}=D_{pp}$ and $D_{np}=D_{pn}$ for symmetric nuclear matter,
whereas all four $D$'s have different values for asymmetric nuclear
matter. The density-gradient coefficients extracted from the MDI
interaction are, however, much larger than the value of $132 \pm 12
$ MeV fm$^5$ shown in the figure that was used by
Oyamatsu~\cite{Oya03} to fit the nuclear radii and is also
consistent with the average value from different Skymre
interactions. We note that the large density-gradient coefficients
in the presence study is due to our assumption that all of the
momentum dependence in the MDI interaction comes from the
finite-range exchange term.

\begin{figure}[h]
\centerline{\includegraphics[scale=0.8]{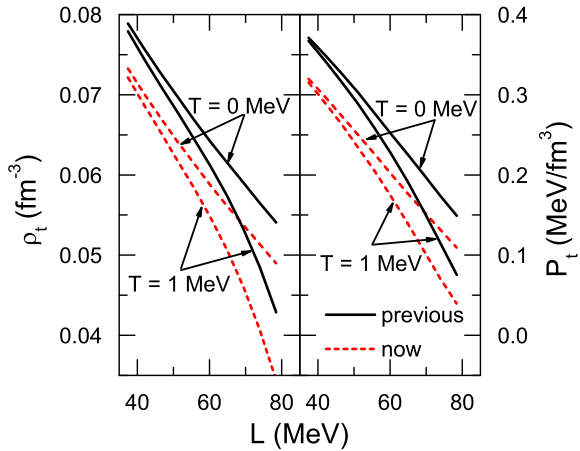}} \caption{(Color
online) The transition density $\rho_t$ and pressure $P_t$ at the
inner edge of neutron star crust as functions of the slope parameter
$L$ for $T=0$ MeV and $T=1$ MeV. Previous
results{~\cite{XCLM09,Xu10}} based on density-gradient terms from
Ref.~\cite{Oya03} are also shown for comparison.} \label{rhotPt}
\end{figure}

In Fig.~\ref{rhotPt}, we compare the transition density and pressure
in this study with those from previous
calculations~\cite{XCLM09,Xu10} that use the value of $132$ MeV
fm$^5$ from Ref.~\cite{Oya03} for the density-gradient coefficients
and neglect the exchange Coulomb interaction. It is seen that the
present results are smaller than previous ones for neutron star
temperatures $T=0$ MeV and $T=1$ MeV. We note that including the
exchange Coulomb term for protons slightly increases $\rho_t$ while
the larger density-gradient coefficients make $\rho_t$ smaller. From
the latest constraint $L = 58 \pm 18 $ MeV on the density slope of
the nuclear symmetry energy, the transition density and pressure are
constrained within $0.050$ fm$^{-3} < \rho_t < 0.071$ fm$^{-3}$ and
$0.12$ MeV/fm$^{3} < P_t < 0.31$ MeV/fm$^{3}$ for $T=0$ MeV, and
$0.038$ fm$^{-3} < \rho_t < 0.070$ fm$^{-3}$ and $0.06$ MeV/fm$^{3}
< P_t < 0.30$ MeV/fm$^{3}$ for $T=1$ MeV in the present study.
Although the transition density is smaller at fixed $L$ compared to
our previous results~\cite{XCLM09}, which leads to a smaller crustal
fraction of the moment of inertia for neutron stars and an even
stricter constraint on the masses and radii of neutron stars, the
upper limit values of $\rho_t$ and $P_t$ are larger because of the
smaller values of $L$. The final constraint on the neutron star
mass-radius (M-R) relation is expected to be similar to that in the
previous work~\cite{XCLM09}.

\section{Summary}
\label{sum}

Assuming that the momentum dependence in the MDI interaction is
entirely due to the finite-range exchange term in the
nucleon-nucleon interaction, we have identified the NN interaction
potential that underlies the MDI interaction, which has been
extensively used in studying isospin effects in intermediate-energy
heavy-ion collisions as well as the properties of hot nuclear matter
and neutron star matter through the resulting potential energy
density in infinite nuclear matter. Using the density matrix
expansion, we have obtained an effective zero-range Skyrme-like
interaction with density-dependent parameters from the finite-range
exchange interaction. Compared to the density-gradient coefficients
extracted by Oyamatsu~\cite{Oya03} and the average value from
different Skyrme interactions, the values from the MDI interaction
are much larger. Since the momentum dependence in the MDI
interaction could also come from the intrinsic momentum dependence
in the elementary nucleon-nucleon interaction or be induced by the
effects of short-range nucleon-nucleon correlations, we have thus
overestimated the effect of the finite-range exchange term. As an
application, the resulting interaction is used to determine the
transition density and pressure at the inner edge of neutron star
crusts based on the linearized Vlasov equation for the neutron star
matter. Fortunately, the instability of neutron star matter is
dominated by the bulk properties of nuclear matter, the large
density-gradient coefficients due to the finite-range interaction do
not reduce much the transition density and pressure in neutron
stars.

\appendix
\section{From the NN interaction to the potential energy density}
\label{appa}

The two particle state $|ij>$  in Eq.~(\ref{vV}) contains the
spatial, spin and isospin parts as
\begin{equation}
|ij>=|i_r i_\sigma i_\tau j_r j_\sigma j_\tau>=|i_r j_r>|i_\sigma
j_\sigma>|i_\tau j_\tau>.
\end{equation}
For the inner product of spatial state, we have
\begin{eqnarray}
&&\sum_{i,j}<i_r j_r|i_r j_r> \notag\\
&=& \sum_{i,j} \int <i_r j_r|\vec{r}_1 \vec{r}_2> <\vec{r}_1
\vec{r}_2 |i_r j_r> d^3 r_1 d^3 r_2
\notag\\
&=& \sum_{i,j} \int \phi_i^\star(\vec{r}_1)\phi_j^\star(\vec{r}_2)
\phi_i(\vec{r}_1)\phi_j(\vec{r}_2) d^3r_1 d^3r_2\notag\\
&=& \int \rho(\vec{r}_1)\rho(\vec{r}_2) d^3 r_1 d^3 r_2
\end{eqnarray}
with $\phi_i(\vec{r})=<\vec{r}|i_r>$ being the space wave function
for particle $i$ and the density $\rho(\vec{r}) =
\sum_i\phi_i^\star(\vec{r})\phi_i(\vec{r})$, and
\begin{eqnarray}
&&\sum_{i,j}<i_r j_r|j_r i_r> \notag\\
&=& \sum_{i,j} \int <i_r j_r|\vec{r}_1 \vec{r}_2> <\vec{r}_1
\vec{r}_2 |j_r i_r> d^3 r_1 d^3 r_2
\notag\\
&=& \sum_{i,j} \int \phi_i^\star(\vec{r}_1)\phi_j^\star(\vec{r}_2)
\phi_j(\vec{r}_1) \phi_i(\vec{r}_2) d^3 r_1 d^3 r_2 \notag\\
&=& \int \rho(\vec{r}_1,\vec{r}_2)\rho(\vec{r}_2,\vec{r}_1) d^3 r_1
d^3 r_2
\end{eqnarray}
with the off-diagonal density $\rho(\vec{r}_1,\vec{r}_2) = \sum_i
\phi_i^\star(\vec{r}_1)\phi_i(\vec{r}_2)$, which reduces to the
density when $\vec{r}_1=\vec{r}_2$.

For the zero-range term $v_0 = \frac{1}{6}t_3(1+x_3
P_\sigma)\rho^\alpha(\frac{\vec{r}_1+\vec{r}_2}{2})
\delta(\vec{r}_1-\vec{r}_2)$ in the MDI interaction, its direct
contribution to the energy can be calculated using the fact that
$P_\sigma=1/2$ when considering the inner product of spin state
in spin saturated matter, and the result is
\begin{eqnarray}
E_0^D &=& \frac{1}{2} \sum_{i,j} <ij|v_0|ij> \notag\\
&=& \frac{t_3}{12} \int d^3 r_1 d^3 r_2 \rho(\vec{r}_1) \rho(\vec{r}_2)
\rho^\alpha\left(\frac{\vec{r}_1+\vec{r}_2}{2}\right)
\delta(\vec{r}_1-\vec{r}_2) \notag\\
&\times&\left(1+\frac{x_3}{2}\right) \notag\\
&=& \int d^3r H_0^D (\vec{r})
\end{eqnarray}
with
\begin{equation}
H_0^D = \frac{t_3}{12} \left(1+\frac{x_3}{2}\right) \rho^{\alpha+2}.
\end{equation}
Its exchange contribution
\begin{equation}
E_0^E = \frac{1}{2} \sum_{i,j} <ij|v_0(-P_rP_\sigma P_\tau) |ij>
\end{equation}
can be evaluated by using $(1+x_3P_\sigma)(-P_\sigma P_\tau) =
-P_\sigma P_\tau - x_3 P_\tau$ and replacing $P_\tau$ by
$\delta_{i_\tau,j_\tau}$ when considering the inner product of
isospin state, and the result is
\begin{eqnarray}
E_0^E &=& -\frac{t_3}{12} \int d^3 r_1 d^3 r_2
\rho(\vec{r}_1,\vec{r}_2) \rho(\vec{r}_2,\vec{r}_1)
\rho^\alpha\left(\frac{\vec{r}_1+\vec{r}_2}{2}\right)
\notag\\
&\times& \delta(\vec{r}_1-\vec{r}_2) \left(\frac{1}{2}+x_3\right) \frac{1+\delta^2}{2}\notag\\
&=& \int d^3r H_0^E (\vec{r})
\end{eqnarray}
with
\begin{equation}
H_0^E = -\frac{t_3}{12} \left(\frac{1}{2}+x_3\right) \rho^{\alpha+2}
\frac{1+\delta^2}{2},
\end{equation}
and $\delta$ is the isospin asymmetry. The total contribution of the
zero-range interaction to the potential energy density is thus
\begin{equation}
H_0 = H_0^D + H_0^E =
\frac{t_3}{16}\rho^{\alpha+2}\left(1-\frac{1+2x_3}{3}\delta^2\right).
\end{equation}
Comparing it with Eq.~(\ref{MDIV}) leads to
Eqs.~(\ref{t3})$-$(\ref{alpha}).

For the contribution from the finite-range interaction in
Eq.~(\ref{MDIv}) to the energy, we use in the exchange term the
relation $(W+BP_\sigma-HP_\tau-MP_\sigma P_\tau)(-P_\sigma P_\tau) =
M+H P_\sigma-BP_\tau-W P_\sigma P_\tau$ to obtain the following
total contribution to the potential energy
\begin{eqnarray}\label{er}
E_r &=& \frac{1}{2} \int d^3 r_1 d^3 r_2 \{[\rho_n(\vec{r}_1)\rho_n(\vec{r}_2)+\rho_p(\vec{r}_1)
\rho_p(\vec{r}_2)]\notag\\
&\times&\left(W+\frac{B}{2}-H-\frac{M}{2}\right)+2\rho_n(\vec{r}_1)\rho_p(\vec{r}_2)
\left(W+\frac{B}{2}\right)\notag\\
&+& [\rho_n(\vec{r}_1,\vec{r}_2)\rho_n(\vec{r}_2,\vec{r}_1)+\rho_p(\vec{r}_1,\vec{r}_2)
\rho_p(\vec{r}_2,\vec{r}_1)]\notag\\
&\times&\left(M+\frac{H}{2}-B-\frac{W}{2}\right) \notag\\
&+& 2\left. \rho_n(\vec{r}_1,\vec{r}_2) \rho_p(\vec{r}_2,\vec{r}_1)\left(M+\frac{H}{2}\right)\right\}\frac{e^{-\mu
|\vec{r}_1-\vec{r}_2}|}{|\vec{r}_1-\vec{r}_2|},
\end{eqnarray}
with the first and second terms being the direct contribution, and
the third and fourth terms being the exchange contribution.

To obtain the energy density functional from the finite-range
interaction, we introduce the coordinate transformation
\begin{eqnarray}\label{trans}
\vec{r} = (\vec{r}_1 + \vec{r}_2)/2,\quad\vec{s} = \vec{r}_1 -
\vec{r}_2.
\end{eqnarray}
For the direct contribution, we use the approximation of infinite
nuclear matter with constant density and make use of the integral
$\int d^3 s e^{-\mu s}/s=4\pi/\mu^2$ to obtain
\begin{equation}
E_r^D = \int d^3 r H_r^D (\vec{r})
\end{equation}
with
\begin{equation}\label{vrd}
H_r^D =
\frac{2\pi}{\mu^2}\left(W+\frac{B}{2}-\frac{H}{2}-\frac{M}{4}\right)\rho^2
-\frac{2\pi}{\mu^2}\left(\frac{H}{2}+\frac{M}{4}\right)\rho^2
\delta^2.
\end{equation}
For the exchange contribution, we express the density matrix in
terms of the Fourier transform of the Wigner function, that is
\begin{equation}\label{wigner}
\rho_\tau\left(\vec{r}+\frac{\vec{s}}{2},\vec{r}-\frac{\vec{s}}{2}\right)
= \int d^3 p f_\tau (\vec{r},\vec{p}) e^{-i\vec{p} \cdot \vec{s}}.
\end{equation}
Substituting Eq.~(\ref{wigner}) into Eq.~(\ref{er}) and evaluating
the Fourier transform of the Yukawa interaction according to
\begin{equation}
\int d^3 s e^{-i(\vec{p}-\vec{p}^\prime) \cdot \vec{s}} \frac{e^{-\mu s}}{s} =
\frac{4\pi}{\mu^2} \frac{1}{1+(\vec{p}-\vec{p}^\prime)^2/\mu^2},
\end{equation}
the exchange contribution of the finite-range interaction to the
potential energy is then
\begin{equation}
E_r^E = \int d^3 r H_r^E(\vec{r}),
\end{equation}
with
\begin{eqnarray}\label{vre}
H_r^E(\vec{r}) &=& \frac{2\pi}{\mu^2} \int \frac{d^3 p d^3
p^\prime}{1+(\vec{p}-\vec{p}^\prime)^2/\mu^2}
\left[\left(M+\frac{H}{2}-B-\frac{W}{2}\right)\right.\notag\\
&\times&(f_n(\vec{r},\vec{p})f_n(\vec{r},\vec{p}^\prime)+f_p(\vec{r},\vec{p})f_p(\vec{r},\vec{p}^\prime))
\notag\\
&+&\left.\left(M+\frac{H}{2}\right)2f_n(\vec{r},\vec{p})f_p(\vec{r},\vec{p}^\prime)\right].
\end{eqnarray}
Comparing Eqs.~(\ref{vrd}) and (\ref{vre}) with Eq.~(\ref{MDIV})
then allows one to obtain Eqs.~(\ref{lambda})$-$(\ref{m}) for the
parameters in the finite-range Yukawa potential in the MDI
interaction.

\section{Density matrix expansion of the finite-range exchange interaction}
\label{appb}

Here we follow exactly the same procedure for the density matrix
expansion introduced in Ref.~\cite{Neg72}. To obtain the energy
density functional in Eq.~(\ref{hfh}) from the exchange contribution
in Eq.~(\ref{er}), we first use the coordinate transformation given
by Eq.~(\ref{trans}) to express the density matrix as
\begin{eqnarray}
&&\rho_\tau(\vec{r}_1,\vec{r}_2)\notag\\
&=& \rho_\tau(\vec{r}+\frac{\vec{s}}{2},\vec{r}-\frac{\vec{s}}{2})\notag\\
&=&\sum_i \phi_{\tau i}^\star(\vec{r}+\frac{\vec{s}}{2})\phi_{\tau i}(\vec{r}-\frac{\vec{s}}{2})\notag\\
&=& e^{ \frac{\vec{s}}{2}\cdot(\nabla_1-\nabla_2)} \sum_i \phi_{\tau
i}^\star(\vec{r})\phi_{\tau i}(\vec{r}),
\end{eqnarray}
with $\nabla_{1(2)}$ acting on the first (second) term on the right,
and $\phi_{\tau i}$ is the spatial wave function of particle $i$
with isospin $\tau$. The angular average over the direction of
$\vec{s}$ is then
\begin{eqnarray}
&&\frac{1}{4\pi} \int d\Omega_s \rho_\tau(\vec{r}+\vec{\frac{s}{2}},\vec{r}-\vec{\frac{s}{2}})\notag\\
&=&
\frac{\sinh[\frac{\vec{s}}{2}\cdot(\nabla_1-\nabla_2)]}{\frac{\vec{s}}{2}\cdot(\nabla_1-\nabla_2)}
\sum_i \phi_{\tau i}^\star(\vec{r})\phi_{\tau i}(\vec{r}).
\end{eqnarray}
Using the Bessel-function expansion
\begin{equation}
\frac{\sinh(xy)}{xy}=\frac{1}{x} \sum_{n=0}^{+\infty} (4n+3) j_{2n+1}(x) Q_n(y^2)
\end{equation}
with $j_{2n+1}$ the $(2n+1)$th-order spherical Bessel function and
\begin{equation}
Q_n(y^2) = \frac{1}{2^{2n+1}} \sum_{l=0}^n
\frac{(4n+2-2l)!y^{2(n-l)}}{l!(2n+1-l)!(2n+1-2l)!},
\end{equation}
one obtains by identifying $x\sim k_\tau s$ and $y \sim
(\nabla_1-\nabla_2)/2k_\tau $ with $k_\tau$ being the Fermi momentum
and keeping up to the third-order Bessel function the following
result:
\begin{eqnarray}
&&\frac{1}{4\pi} \int d\Omega_s \rho_\tau(\vec{r}+\vec{\frac{s}{2}},\vec{r}-\vec{\frac{s}{2}})\notag\\
&=& \rho_{SL}^{}(k_\tau s) \rho_\tau(\vec{r}) + g(k_\tau s) s^2 \notag\\
&\times&\left[\frac{1}{4}\nabla^2
\rho_\tau(\vec{r})-\tau_\tau(\vec{r}) +\frac{3}{5} k_\tau^2
\rho_\tau(\vec{r})\right]
\end{eqnarray}
with $\rho_{SL}^{}(k_\tau s) = \frac{3}{k_\tau s}j_1(k_\tau s)$,
$g(k_\tau s)=\frac{35}{2(k_\tau s)^3}j_3(k_\tau s)$ and
$\tau_\tau(\vec{r})=\sum_i |\nabla \phi_{\tau i}(\vec{r})|^2$
being the kinetic energy density. By making the
approximation
\begin{eqnarray}\label{exrho}
&&\frac{1}{4\pi} \int d\Omega_s
\rho_\tau(\vec{r}+\vec{\frac{s}{2}},\vec{r}-\vec{\frac{s}{2}})
\rho_{\tau^\prime}(\vec{r}-\vec{\frac{s}{2}},\vec{r}+\vec{\frac{s}{2}})\notag\\
&\approx& \rho_\tau(\vec{r})\rho_{SL}^{}(k_\tau s)\rho_{\tau^\prime}(\vec{r})\rho_{SL}^{}(k_{\tau^\prime} s) \notag\\
&+& \rho_\tau(\vec{r})\rho_{SL}^{}(k_\tau s)g(k_{\tau^\prime} s)s^2\notag\\
&\times&\left[\frac{1}{4}\nabla^2 \rho_{\tau^\prime}(\vec{r})-\tau_{\tau^\prime}(\vec{r})+\frac{3}{5} k_{\tau^\prime}^2 \rho_{\tau^\prime}(\vec{r})\right] \notag\\
&+& \rho_{\tau^\prime}(\vec{r})\rho_{SL}^{}(k_{\tau^\prime} s)g(k_\tau s)s^2\notag\\
&\times&\left[\frac{1}{4}\nabla^2
\rho_\tau(\vec{r})-\tau_\tau(\vec{r})+\frac{3}{5}
k_\tau^2\rho_\tau(\vec{r})\right],
\end{eqnarray}
and using the relation
\begin{eqnarray}\label{fdrho}
F(\rho_n,\rho_p) \nabla^2 \rho_n &=& -\frac{\partial F(\rho_n,\rho_p)}{\rho_n}(\nabla \rho_n)^2 \notag\\
&-&\frac{\partial F(\rho_n,\rho_p)}{\rho_p}\nabla \rho_n \cdot
\nabla\rho_p.
\end{eqnarray}
we obtain
\begin{equation}
E_r^E = \int d^3r H_{SL}^E (\vec{r}),
\end{equation}
where $H_{SL}^E(\vec{r})$ is the Skymre-like zero-range energy
density from the finite-range exchange interaction defined in
Eq.~(\ref{hfh}).

\begin{acknowledgments}

We thank Lie-Wen Chen for helpful discussions. This work was
supported in part by the U.S. National Science Foundation under
Grant No. PHY-0758115 and the Welch Foundation under Grant No.
A-1358.

\end{acknowledgments}

\end{document}